\documentclass[aps,twocolumn,nofootinbib,prd,eqsecnum,showpacs,showkeys,preprintnumbers]{revtex4-1}
\usepackage[caption=false]{subfig}
\usepackage{graphicx}
\usepackage{amsmath}
\usepackage{amsfonts}
\usepackage{amssymb}
\usepackage{color}
\usepackage{bm}
\usepackage{mathrsfs}
\usepackage{epstopdf}
\usepackage{url}
\usepackage{footnote}
\usepackage{textcomp}
\usepackage{ulem}
\usepackage{esint}
\usepackage[unicode=true, pdfusetitle,
 bookmarks=true,bookmarksnumbered=false,bookmarksopen=false,
 breaklinks=false,pdfborder={0 0 1},backref=false,colorlinks=false]{hyperref}
\usepackage{multirow}
\usepackage{pifont}
\DeclareMathOperator{\arcsinh}{arcsinh}
\begin{document}
\title{Non-minimal Higgs inflation within holographic cosmology}

\author{Aatifa Bargach$^{1}$}
\email{a.bargach@ump.ac.ma}
\author{Farida Bargach$^{1}$}
\email{f.bargach@ump.ac.ma}
\author{Mariam Bouhmadi-L\'{o}pez$^{2,3}$}
\email{mariam.bouhmadi@ehu.eus}
\author{Taoufik Ouali$^{1}$}
\email{ouali\_ta@yahoo.fr}
\affiliation{$^{1}$Laboratory of Physics of Matter and Radiation, \\
University of Mohammed first, BP 717, Oujda, Morocco\\
$^{2}$ Department of Theoretical Physics University of the Basque Country UPV/EHU. P.O. Box 644, 48080 Bilbao, Spain\\
$^{3}$ IKERBASQUE, Basque Foundation for Science, 48011, Bilbao, Spain}
\date{\today }
\begin{abstract}
We derive a Higgs inflationary model in the context of holographic cosmology, where we consider a universe filled with a Higgs field non-minimally coupled to gravity in a slow-roll regime. The amplitude of density (scalar) perturbations is calculated. In this regard, we show that the background and perturbative parameters characterising the inflationary era are related to the standard one through corrections terms. We found that for the e-fold number $N\sim 58$, the spectral index,$n_{r}$, and the tensor-to-scalar ratio,$r$, values are $0.965$ and $0.021$, respectively, which are in agreement with 2018 Planck observational data. However, as soon as we move from $N\sim 58$, the model is ruled out by the current data. 
\end{abstract}
\maketitle
\section{Introduction}
Inflation, which is a phase of accelerated expansion in the early universe, is originally aimed to  solve some of the hot big bang shortcomings such as the flatness, the horizon and the primordial monopole problems \cite{Starobinsky:1980te, Guth:1980zm, Linde:1984st}. The inflationary scenario is not only appealing for solving these problems, but also provides solutions to the primordial density fluctuations which are necessary for the formation of the large scale structure observed in the present universe \cite{Liddle:2000cg}. Indeed, the inhomogeneities originated from the quantum vacuum fluctuations of the inflaton field are stretched on large scales, through the accelerated expansion, and become classical to give rise to the structures we observe nowadays. The theory of cosmological perturbations is the key to study the inflationary scenario and it provides also a good framework to determine the cosmic microwave background (CMB) prediction as well as to compute the power spectrum  in order to make connection with the observational data \cite{Akrami:2018odb}.\par

Despite its great successes, the inflationary paradigm suffers from some problems especially at high enough energies \cite{Lidsey:1995np,Riotto:2002yw, Brandenberger:2005nz}, where general relativity (GR) should be modified. Within this spirit modification of gravity has been the base of several models that have attracted so much attention in the recent years, such as the braneworld scenario \cite{Maartens:1999hf, Zhang:2004in, delCampo:2007cia} and models with a non-minimally coupled (NMC) inflaton field \cite{Salopek:1988qh, Fakir:1990eg, Faraoni:1996rf, Faraoni:2000wk}.\par
In the braneworld scenario, matter is confined on the 3-brane, while gravity can propagate in the $5$-dimensional bulk. One of the most interesting braneworld models was proposed by Randall and Sundrum (RS2) \cite{Randall:1999vf}, where our universe corresponds to a $4$-dimensional single brane embedded in a bulk corresponding to pieces of a $5$-dimensional anti-de Sitter space-time (AdS$_{5}$). This model provides also a good framework for exploring holographic ideas that have emerged in M-theory. The AdS/CFT correspondence has been thoroughly studied in RS2 model and, in particular, the cosmology of a homogeneous and isotropic brane whithin this framework \cite{Kiritsis:2005bm, Bilic:2015uol}. On its original formulation, the AdS/CFT correspondence suggests that the $5$-dimensional gravitational dynamics may be determined from the (quantum) dynamics of the fields on a lower-dimensional boundary.\par
Among the many models of inflation, the simplest realisation comprise the introduction of a scalar {\color{black} field} (the inflaton). This hypothetical particle may appear in different extensions of the standard model of elementary particles. However, the only scalar field that has been detected so far is the Higgs boson, whose existence was confirmed in $2013$ at the LHC \cite{Aad:2012tfa,Chatrchyan:2012xdj}. The idea that the Higgs field plays the role of the inflaton field has been already discussed, for example in \cite{Rehman:2010es,Bezrukov:2007ep,CervantesCota:1995tz,Bezrukov:2019ylq,Raatikainen:2019qey,Lee:2019efp}. This is an excellent way to connect inflationary models with known particle physics and benefit from observational constraints of both particle physics and cosmology, this way the model will be more predictive.\par

Nevertheless, the major problem with Higgs inflation is that the energy scale of the Higgs field is too small to generate enough e-folds required to solve the aforementioned problems of the big bang cosmology \cite{Isidori:2007vm}. Indeed, the Higgs self-coupling value leading to successful inflation, $\lambda \sim 10^{-13}$, is not compatible with the required value suggested by the standard model from the measured Higgs bosons mass $m=125 \; GeV$, $\lambda \sim 0.13$. By postulating a NMC between the Higgs field and the Ricci scalar we ensure an excellent agreement with observational data \cite{Akrami:2018odb}. Early ideas to consider a NMC with the Higgs field were formulated, for example in \cite{Bezrukov:2007ep}.\par

The assumption of considering a NMC inflaton field to gravity was already introduced quite some years ago \cite{Salopek:1988qh, Fakir:1990eg}, where the motivation  arises at the quantum level when quantum corrections to the scalar field theory are considered. Recently, this identification of a NMC term has became again popular \cite{Chiba:2008ia, Tamanini:2010uq, Markkanen:2014dba, Racioppi:2017spw, Shojaee:2018fre,Azevedo:2018nvi,Wang:2019spw, Shokri:2019rfi,Bargach:2018ujh,Savchenko:2018pdr,Tang:2019jkn,Novello:2019hxu}, supported by CMB measurements as provided by the Planck Satellite \cite{Akrami:2018odb} as well as by the discovery of the Higgs boson \cite{Aad:2012tfa,Chatrchyan:2012xdj}. Several studies in this field with some constraints on the value of the NMC constant have being carried in order to have a theoretically consistent picture of the inflationary model \cite{Spokoiny:1984bd, Futamase:1987ua, Kaiser:1994vs,Faraoni:1997fn, Chiba:2008ia, Tamanini:2010uq, Markkanen:2014dba, Racioppi:2017spw, Shojaee:2018fre,Azevedo:2018nvi,Wang:2019spw,Shokri:2019rfi}. In the case of NMC Higgs inflation, it is necessary to assume a large coupling constant to improve the situation with the Higgs self-coupling value. For the NMC Higgs inflation model \cite{Barvinsky:2008ia,Bezrukov:2007ep, DeSimone:2008ei, Barvinsky:2009fy, Barvinsky:2009ii, Bezrukov:2010jz, Bezrukov:2013fka} the NMC constant, $\alpha_{0}$, is of order of $\sim 10^{4}$, while an upper bound required to be of order of $\alpha_{0} \ll 2.6\times 10^{15}$ was derived in \cite{Atkins:2012yn}.\par

Furthermore, holographic cosmology seems a good framework to study constraints on inflationary parameters. Indeed, it was found that the AdS/CFT duality may describe the inflationary era and provide good agreement with observational constraints for a universe filled with a scalar field \cite{Lidsey:2005nt} or a tachyon field \cite{Bouabdallaoui:2016izz} or even an induced gravity model for both kind of fields \cite{Bargach:2019pst}.\par 

The purpose of this paper is to analyse the effect of the  AdS/CFT holographic duality on the dynamics of the NMC Higgs field which plays the inflaton role. Indeed, We will consider constraints from current Planck data under the assumption of a NMC Higgs inflaton with a quartic Higgs-like potential in the slow-roll approximation. Within this framework, we will study the cosmological perturbations to obtain the spectral index and the tensor-to-scalar ratio.  We will constrain the model using current observational data \cite{Akrami:2018odb}. In fact, current constraints from Planck data \cite{Akrami:2018odb} suggest an upper limit of the tensor-to-scalar ratio $r < 0.1$ (Planck alone) at 95\% confidence level (CL) and a value of the spectral index $n_{s}=0.9649\pm 0.0042$ quoted to $68$\% CL.\par

This paper is organized as follows. In section \ref{sec1}, we describe the basic setup of NMC to gravity. In section \ref{sec2}, we present the scalar perturbations from an holographic point of view. In section \ref{sec3}, a Higgs inflation model is considered taking into account observational constraints to check the viability of the model. Finally, we present our conclusions in section \ref{sec4}.

\section{Non-minimal coupling to gravity and inflation} \label{sec1}
We consider a generalized Randall Sundrum model with a NMC scalar field localised on the brane whose action reads \cite{BouhmadiLopez:2004ys,Nozari:2012cy,Bargach:2018ujh}
\begin{widetext}
\begin{equation}\label{action}
S=\int_{bulk}d^{5}x\sqrt{-g^{\left( 5\right) }}\left( \frac{1}{2\kappa_{5}^{2}}R_{5}-\Lambda
_{5}\right)+\int_{brane}d^{4}x\sqrt{-\overset{}{g}}
\left(f(\phi) R-\frac{1}{2} g^{\mu \nu }\nabla _{\mu }\phi \nabla _{\upsilon }\phi -V(\phi)-\Lambda_{4}\right),
\end{equation}
\end{widetext}
{\color{black}where $\kappa_{5}^{2}$, $R_{5}$ and $\Lambda_{5}$ are the 5D gravitational constant, the Ricci scalar of the $5$-dimensional metric $g^{(5)}$ and the bulk cosmological constant, respectively}. In the brane action, $R$ is the Ricci scalar of the induced metric $g$, $f(\phi)$ is the NMC of the scalar field $\phi$ to induced gravity on the brane, $V(\phi )$ is the scalar field potential and $\Lambda_{4}$ is the brane tension.\par

The gravitational equations on the brane, for a vanishing cosmological constant, can be obtained from Eq. \eqref{action} as \cite{BouhmadiLopez:2004ys}
\begin{equation}\label{Gmodifie}
  G^{(4)}_{\mu\nu}= \kappa_{4}^{2} \tilde{T}_{\mu\nu} +\kappa_{5}^{4} \Pi_{\mu\nu}- E_{\mu\nu},
\end{equation}
where $\kappa_{4}^{2}$ is related to the gravitaional constant on the brane, the total energy-momentum tensor $\tilde{T}_{\mu\nu}$ is given by
\begin{align}\label{Tmunutilde}
\tilde{T}_{\mu\nu}=& T_{\mu\nu} -2 f(\phi) G_{\mu\nu}^{(4)},\\
\label{Tmunu}
  T_{\mu\nu} =&\nabla_{\mu}\phi \nabla_{\nu}\phi-\frac{1}{2}g_{\mu\nu} (\nabla\phi)^{2}-g_{\mu\nu} V(\phi)
 \nonumber \\
  &+2 \nabla_{\mu} \nabla_{\nu}f -2\Box f g_{\mu\nu},
\end{align}
the quadratic energy momentum tensor $\Pi_{\mu\nu}$ reads
\begin{equation}\label{Pimunu}
  \Pi_{\mu\nu}= -\frac{1}{4} \tilde{T}^{\lambda}_{\mu} \tilde{T}_{\lambda\nu} + \frac{1}{12} \tilde{T} \tilde{T}_{\mu\nu} +\frac{1}{8} g_{\mu\nu} (\tilde{T}^{\alpha\beta} \tilde{T}_{\alpha\beta}-\frac{1}{3} \tilde{T}^{2}),
\end{equation}
and $E_{\mu\nu}$ is the projected Weyl tensor on the brane which represents the effect of the bulk geometry. \par
The total energy-momentum tensor is conserved on
the brane as \cite{BouhmadiLopez:2004ys}
\begin{equation}\label{continuite}
\nabla_{\mu}\tilde{T}^{\nu}_{\mu}=0.
\end{equation}

The equation of motion takes the following form \cite{BouhmadiLopez:2004ys}
\begin{equation}\label{motion}
 \ddot{\phi}+3H\dot{\phi}+V_{,\phi}-f_{,\phi} R=0,
 \end{equation}
where {\color{black}the} dot corresponds to a derivative with respect to the cosmic time and {\color{black}the} subscript ($_{,\phi}$) denotes a derivative with respect to {\color{black}the} scalar field $\phi$ and \mbox{$R=6 (\dot{H}+2H^{2})$}.\par
On the other hand, the gravitational field equations of the brane, for a vanishing cosmological constant, within the holographic scope can be written as \cite{Kanno:2002iaa,Lidsey:2005nt,Bargach:2018ujh}
\begin{equation}\label{Gmod}
   G^{(4)}_{\mu\nu}= \kappa_{4}^{2} \left( \tilde{T}_{\mu\nu} + T_{\mu\nu}^{CFT} \right),
\end{equation}
where $\tilde{T}_{\mu\nu}$ is the total energy-momentum tensor given in Eq. \eqref{Tmunutilde}, and $T_{\mu\nu}^{CFT}$ denotes the energy-momentum tensor of the cutoff version of conformal field theory (see \cite{Lidsey:2005nt} for further details). We highlight that Eq. \eqref{Gmod} can be rewritten as Eq. \eqref{Gmodifie} (see Ref. \cite{Kanno:2002iaa}). We will use this fact when calculating the perturbations. \par
The {\color{black}modified} Friedmann equation {\color{black}in the context of the holographic view point} for a homogeneous and isotropic universe  with a vanishing spatial curvature
and in absence of a cosmological constant reads \cite{Bargach:2018ujh}
\begin{equation}\label{freidmanmod}
 H^{2} = \frac{1}{4 c \kappa_{\textrm{ eff}}^{2}} \left [ 1\pm \sqrt{1- \frac{\rho}{\rho_{max}}}\right],
\end{equation}
where $c$ is the conformal anomaly coefficient, \mbox{$\kappa_{\textrm{ eff}}^{2}=\kappa_{4}^{2}/(1+2 \kappa_{4}^{2}f(\phi))$}, $\rho$ is the energy density of the brane, \mbox{$\rho_{max} = 3/8c \kappa_{\textrm{ eff}}^{4}$} and the sign ($\pm$) shows the existence of two branches of solutions. We recover the standard form of the Friedmann equation at low-energy limit $\rho\ll \rho_{max}$ and for the limit $f\rightarrow 0$, but only on the negative branch. On what follows, we will focus our analysis on this branch. From Eq. \eqref{Tmunu}, the energy
density and pressure can be defined as
\begin{align}
 \rho &=\frac{1}{2}\dot{\phi}^{2}+V-6 \dot{f} H,\label{rho}\\
 p&=\frac{1}{2}\dot{\phi}^{2}-V+2(\ddot{f}+2H\dot{f}).\label{p}
\end{align}

During the inflationary epoch and assuming a slow-roll expansion \cite{Torres:1996fr}, i.e. $\dot{\phi}^{2}\ll V$, $\ddot{\phi}\ll 3H\dot{\phi}$ and $\ddot{f}\ll H\dot{f}\ll H^{2} f$ the Friedmann equation
\eqref{freidmanmod} reduces to
\begin{equation}\label{friedmanslow}
  H^{2}\simeq \frac{1}{4 c \kappa_{\textrm{ eff}}^{2}} \left [ 1- \sqrt{1- U}\right],
\end{equation}
where the dimensionless parameter $U$ is given by
\begin{equation}\label{U}
U\equiv \frac{V}{V_{max}}= \frac{8 c \kappa_{\textrm{eff}}^{4} V}{3}.
\end{equation}
Also, the equation of motion \eqref{motion} reduces to
\begin{equation}\label{motionslow}
  \dot{\phi}\simeq-\frac{V_{,\phi}-f_{,\phi} R}{3H}.
\end{equation}

\section{Scalar perturbations from holographic cosmology} \label{sec2}
In this section, we analyse  the generation of the cosmological perturbations from the seeds corresponding to the quantum fluctuations taking place during  inflation. Scalar perturbations of a Friedmann-Lemaître-Robertson-Walker (FLRW) background in the conformal Newtonian gauge are given by \cite{Bardeen:1980kt,Mukhanov:1990me}
\begin{equation}\label{metric}
  ds^{2}=-(1+2\Phi)dt^{2}+a^{2}(t)(1-2\Psi)\delta_{ij}dx^{i}dx^{j},
\end{equation}
where $a(t)$ is the scale factor, $\Phi(t, x)$ and 	$\Psi(t, x)$ are the scalar perturbations. Our starting point is the perturbed Einstein equations \cite{Deffayet:2002fn}
\begin{equation}\label{perturbedG}
  \delta G^{(4)\mu}_{\quad \;\nu}= \kappa_{4}^{2} \delta \tilde{T}_{\;\nu}^{\mu} +\kappa_{5}^{4} \delta\Pi_{\;\nu}^{\mu}- \delta E_{\;\nu}^{\mu}.
\end{equation}
For the above perturbed metric, one can obtain the individual components of Eq. \eqref{perturbedG} as follows \cite{Deffayet:2002fn}
\begin{subequations}\label{Gperturbed}
\begin{widetext}
\begin{align}
6 H(\dot{\Psi}+H\Phi)-2\frac{\nabla^{2}}{a^{2}}\Psi=&\kappa_{4}^{2}\delta \tilde{T}_{\;0}^{0} +\kappa_{5}^{4} \delta\Pi_{\;0}^{0}- \delta E_{\;0}^{0} , \\
  -2(\dot{\Psi}+H\Phi)_{,i} =&\kappa_{4}^{2}\delta \tilde{T}_{\;i}^{0} +\kappa_{5}^{4} \delta\Pi_{\;i}^{0}- \delta E_{\;i}^{0}, \\
6 \ddot{\Psi}+6(3H^{2}+2\dot{H})\Phi+6H(\dot{\Phi}+3\dot{\Psi})+4\frac{\nabla^{2}}{a^{2}}(\Phi-\Psi) =&\kappa_{4}^{2}\delta \tilde{T}_{\;i}^{i} +\kappa_{5}^{4}
\delta\Pi_{\;i}^{i}- \delta E_{\;i}^{i},  \\
   \frac{1}{a^{2}}(\Psi-\Phi)_{\;,j}^{,i}=&\kappa_{4}^{2}\delta \tilde{T}_{\;j}^{i} +\kappa_{5}^{4} \delta\Pi_{\;j}^{i}- \delta E_{\;j}^{i}, \qquad i\neq j.
\end{align}
\end{widetext}
\end{subequations}
The right hand side of Eq. \eqref{perturbedG} is the sum of the perturbed energy momentum tensor $\delta\tilde{ T}_{\;\nu}^{\mu}$ given by \cite{Deffayet:2002fn}
\begin{equation}\label{deltaT}
  \delta \tilde{T}_{\;\nu}^{\mu} = \left(
                               \begin{array}{cc}
                                 -\delta\tilde{\rho} & a \delta\tilde{ q_{,i}} \\
                                 -a^{-1} \delta\tilde{ q}^{,i} & \delta\tilde{ p }\delta_{\; j}^{i}+ \delta\tilde{\pi}_{\; j}^{i} \\
                               \end{array}
                             \right),
\end{equation}
where $\delta\tilde{\rho}$ is the perturbed total energy density,  $\delta\tilde{ q}$ is the perturbed total momentum, $\delta\tilde{ p}$ is the perturbed total pressure and the total anisotropic stress tensor is  $\delta\tilde{\pi}^{i}_{j}=(\Delta^{i}_{j}-\frac{1}{3}\delta^{i}_{ j}\Delta)\delta\tilde{\pi}$ with $\Delta^{i}_{j}=\delta^{i}_{ k} \partial_{k}\partial_{j}$ so that one has $\Delta=\Delta_{i}^{i}$. {\color{black}The second part of Eq. \eqref{perturbedG} is} the perturbed quadratic energy momentum tensor $\delta\Pi_{\;\nu}^{\mu}$ {\color{black} which} can be written as \cite{Deffayet:2002fn}
\begin{widetext}
\begin{equation}\label{deltaPi}
  \delta \Pi_{\;\nu}^{\mu} = \left(
                               \begin{array}{cc}
                                 -\frac{1}{6} \tilde{\rho} \delta\tilde{\rho} & \frac{1}{6} a \tilde{\rho} \delta\tilde{ q}_{,i} \\
                                 -\frac{1}{6} a^{-1} \tilde{\rho} \delta\tilde{ q}^{,i} &\qquad \frac{1}{6}\left[(\tilde{\rho}+\tilde{p})\delta\tilde{ \rho}+\tilde{\rho} \delta\tilde{ p}\right]  \delta_{\; j}^{i}- \frac{1}{12} (\tilde{\rho}+3\tilde{p})\delta\tilde{\pi}_{\; j}^{i} \\
                               \end{array}
                             \right),
\end{equation}
\end{widetext}
and the projected Weyl tensor $\delta E_{\;\nu}^{\mu}$. As stated before this tensor is determined by the effect of the bulk geometry, i.e. it cannot be written in the local covariant form. Hence the set of equations \eqref{Gperturbed} is not closed. We may parameterize the scalar perturbations of $E_{\mu\nu}$ as an effective fluid as \cite{Langlois:2000iu, Koyama:2005kd}
\begin{equation}\label{deltaE}
  \delta E_{\;\nu}^{\mu} =-\kappa_{4}^{2} \left(
                               \begin{array}{cc}
                                 -\delta\rho_{E} & a \delta q_{E,i} \\
                                 -a^{-1} \delta q_{E}^{,i} & \quad \frac{1}{3}\delta \rho_{E} \delta_{\; j}^{i}+ \delta\pi_{E\; j}^{i} \\
                               \end{array}
                             \right),
\end{equation}
with $\delta\rho_{E}$ is the density perturbation, $\frac{1}{3}\delta \rho_{E}$ is the isotropic pressure perturbation, $\delta\pi_{E}$
is the anisotropic stress perturbation and $\delta q_{E}$ is the energy flux perturbation (see Ref. \cite{Langlois:2000iu, Koyama:2005kd}). As we have already mentioned, brane parameters cannot be determined freely since they are influenced by the bulk through the boundary conditions \cite{Shtanov:2007dh}. Here, we use the quasi-static approximation which is useful for structure formation. Therefore, we neglect time-derivatives terms relative to gradient terms. It can be shown that $\delta q_{E}=0$ (see the appendix for details and justification of this assumption).\par

We apply a Fourier {\color{black}transformation to the scalar perturbations} in order to study the evolution of the linear perturbations, thus we decompose the function $\psi(t, x)$ into its Fourier components  $\psi_{k}(t)$ as follows
\begin{equation}\label{fourier}
  \psi(t,x)=\frac{1}{(2\pi)^{3/2}}\int e^{-ikx}\psi_{k}(t)d^{3}k,
\end{equation}
where $k$ represents the wave number $k$. 
 The perturbed equations \eqref{Gperturbed} read
\begin{widetext}
\begin{subequations}\label{Gperturbeddd}
\begin{align}
 H(\dot{\Psi}+H\Phi)+\frac{k^{2}}{3a^{2}}\Psi=& -\frac{\kappa_{4}^{2}}{6}\left[\delta\tilde{\rho}\left(1+\frac{\tilde{\rho}}{\sigma}\right)+\delta\rho_{E}\right], \\
  (\dot{\Psi}+H\Phi) =&-\frac{\kappa_{4}^{2}}{2}\left(1+\frac{\tilde{\rho}}{\sigma}\right)a\delta\tilde{ q},\label{0iGperturbeddd} \\
 \ddot{\Psi}+(3H^{2}+2\dot{H})\Phi+H(\dot{\Phi}+3\dot{\Psi})-\frac{2k^{2}}{3a^{2}}(\Phi-\Psi) =&\frac{\kappa_{4}^{2}}{2}\left[\delta\tilde{
 p}\left(1+\frac{\tilde{\rho}}{\sigma}\right)+\delta\tilde{ \rho}\left(\frac{\tilde{\rho}+\tilde{p}}{\sigma}\right)+\frac{2}{3}k^{2}\delta\pi_{E}\right],  \\
   (\Psi-\Phi)=&-\kappa_{4}^{2}a^{2}\left[\delta\tilde{\pi}\left(1-\frac{\tilde{\rho}+3\tilde{p}}{2\sigma}\right)+\delta\pi_{E}\right],
\end{align}
\end{subequations}
\end{widetext}
where $\sigma\equiv \frac{6\kappa_{4}^{2}}{\kappa_{5}^{4}}=\frac{\rho_{max}}{(1+2\kappa_{4}^{2}f)^{2}}$  and we can split the perturbed effective density, the energy flux, the pressure and the anisotropic stress perturbation, respectively, as follows
\begin{widetext}
\begin{subequations}\label{deltarho}
\begin{align}
 \delta\tilde{\rho}=& \delta\rho-6 H^{2}\delta f+12 f \left[H(\dot{\Psi}+ H \Phi)+\frac{k^{2}}{3a^{2}}\Psi\right],\\ \label{deltaqtilde}
 a\delta\tilde{ q}=&a\delta q+4 f \left(\dot{\Psi}+H\Phi\right),\\
 \delta\tilde{ p}=&\delta p+2(3H^{2}+2\dot{H})\delta f-4 f \left[\ddot{\Psi}+(3H^{2}+2\dot{H})\Phi+H(\dot{\Phi}+3\dot{\Psi})-\frac{k^{2}}{3a^{2}}(\Phi-\Psi)\right],\\
 \delta\tilde{ \pi}^{i}_{j}=&\delta\pi^{i}_{j} +2 f \frac{(\Psi-\Phi)^{i}_{j}}{a^{2}}.
\end{align}
\end{subequations}
\end{widetext}
Then the perturbation equations \eqref{Gperturbeddd} reduce to
\newpage
{\small
\begin{widetext}
\begin{subequations} \label{perturbedmodif}
\begin{align}
 H(\dot{\Psi}+H\Phi)+\frac{k^{2}}{3a^{2}}\Psi=&-\frac{\tilde{\kappa}_{\textrm{ eff}}^{2}}{6}\left[\delta\rho+\frac{1}{\left(1+\frac{\tilde{\rho}}{\sigma}\right)}\delta\rho_{E}\right], \\
  (\dot{\Psi}+H\Phi) =&-\frac{\tilde{\kappa}_{\textrm{ eff}}^{2}}{2}a\delta q, \label{perturbedmodif0i}\\
 \ddot{\Psi}+(3H^{2}+2\dot{H})\Phi+H(\dot{\Phi}+3\dot{\Psi})-\frac{2k^{2}}{3a^{2}}(\Phi-\Psi) =&\frac{\tilde{\kappa}_{\textrm{ \textrm{ eff}}}^{2}}{2}
 \left[\delta p + \frac{\tilde{\kappa}_{\textrm{ eff}}^{2}\left(\frac{\tilde{\rho}+\tilde{p}}{\sigma}\right)}{\left(1+\frac{\tilde{\rho}}{\sigma}\right)^{2}}\delta \rho +\frac{2k^{2}}{\left(1+\frac{\tilde{\rho}}{\sigma}\right)}\left(\frac{1}{3}-\frac{2\tilde{\kappa}_{\textrm{ eff}}^{2} \kappa_{4}^{2} f \left( \frac{\tilde{\rho}+\tilde{p}}{\sigma}\right)}{\left(1+\frac{\tilde{\rho}}{\sigma}\right)}\right)\delta\pi_{E}\right],  \\
   (\Psi-\Phi)=&-a^{2}\frac{\tilde{\kappa}_{\textrm{ eff}}^{2}\left[1+2 \kappa_{4}^{2} f (1+\frac{\tilde{\rho}}{\sigma})\right]}{\left(1+\frac{\tilde{\rho}}{\sigma}\right)\left[1+2 \kappa_{4}^{2} f (1-\frac{\tilde{\rho}+\tilde{p}}{2\sigma})\right]}\left(\delta\pi+\delta\pi_{E}\right),
\end{align}
\end{subequations}
\end{widetext}
}
 where $\tilde{\kappa}_{\textrm{ eff}}^{2}=\frac{\kappa_{4}^{2}\left(1+\frac{\tilde{\rho}}{\sigma}\right) }{\left[1+2 \kappa_{4}^{2} f (1+\frac{\tilde{\rho}}{\sigma})\right]}$ and
\begin{widetext}
\begin{subequations}\label{perturbedtensormomentum}
\begin{align}
 \delta \rho =&\left[ -\dot{\phi}^{2}+12 H \dot{f}\right]\Phi+V_{,\phi} \delta\phi+\dot{\phi} \dot{\delta\phi} -6H \dot{\delta f}+\frac{2}{a^{2}}\Delta \delta f+6\dot{f}\dot{\Psi},\\
a\delta q=& -\dot{\phi}\delta\phi+2 H \delta f-2\dot{\delta f}+2 \dot{f} \Phi,\label{perturbedtensormomentum0i} \\
\delta p=& \left[ -\dot{\phi}^{2}-8 \dot{f}H-4\ddot{f}\right]\Phi-V_{,\phi}\delta\phi+\dot{\phi}\dot{\delta\phi}+4H\dot{\delta f}+2\ddot{\delta f} -\frac{4}{3a^{2}}\Delta \delta f-2\dot{f} \left(\dot{\Phi}+2\dot{\Psi}\right),\\
\delta \pi^{i}_{j}=&\frac{2}{a^{2}}\delta f^{i}_{j}.
\end{align}
\end{subequations}
\end{widetext}

The perturbed equation of motion for the scalar field is given by \cite{Nozari:2012cy}
\begin{widetext}
\begin{multline}\label{perturbedmotion}
\ddot{\delta\phi}+3H\dot{\delta\phi}+\left(V_{,\phi\phi}+\frac{k^{2}}{a^{2}}-R f_{,\phi\phi}\right)\delta\phi=\dot{\phi}(3\dot{\Psi}+\dot{\Phi})+ \left(2 f_{,\phi}R-2V_{,\phi}\right)\Phi\\
+2 f_{,\phi}\left[ (\frac{k^{2}}{a^{2}}-3 \dot{H})\Phi -2 \frac{k^{2}}{a^{2}} \Psi -3 \left(\ddot{\Psi}+4 H \dot{\Psi}+H\dot{\Phi}+\dot{H}\Phi+4 H^{2}\Phi\right)\right].
\end{multline}
\end{widetext}
This equation can be strongly simplified within the slow-roll approximation at large scales, $k\ll aH$. This so, as the scales of cosmological interest have spent most of their time far outside the Hubble radius and have re-entered only relatively recently in the universe. Therefore, large scales $k\ll aH$ is an acceptable assumption. When this condition is satisfied $\dot{\Phi}$, $\dot{\Psi}$, $\ddot{\Phi}$ and $\ddot{\Psi}$ can be neglected \cite{Amendola:2005cr, Amendola:2007ni}. Then, Eq. \eqref{perturbedmotion} reduces to
\begin{equation}\label{perturbedmotionslow}
 \ddot{\delta\phi}+3H\dot{\delta\phi}+\left(V_{,\phi\phi}-12 H^{2} f_{,\phi\phi}\right)\delta\phi= -2V_{,\phi}\Phi.
\end{equation}
By following a similar reasoning, using Eqs. \eqref{0iGperturbeddd},  \eqref{deltaqtilde} and \eqref{perturbedtensormomentum0i}, we can therefore relate the scalar perturbation $\Phi$ to the fluctuation of the scalar field $\delta\phi$ as
\begin{equation}\label{Phi}
  \Phi = \frac{\tilde{\kappa}_{\textrm{ eff}}^{2}\left(\dot{\phi}-2Hf_{,\phi}\right)}{2H} \delta\phi.
\end{equation}

Here, we define one of the most commonly used gauge invariant combinations in terms of matter and metric perturbations which is the comoving curvature perturbation given by \cite{Bassett:2005xm}
\begin{equation}\label{curvature}
\mathcal{R}\equiv \Psi-\frac{H}{\tilde{\rho}+\tilde{P}}\delta\tilde{ q}.
\end{equation}
From Eqs. \eqref{rho}, \eqref{p} and \eqref{0iGperturbeddd} by adopting the slow-roll {\color{black}regime} and the large scale condition, we find
\begin{equation}\label{cur}
 \mathcal{R}=\Psi+\frac{H}{\dot{\phi}\left[1+2 \kappa_{4}^{2}f\left(1+\frac{\tilde{\rho}}{\sigma} \right)\right]} \delta\phi.
\end{equation}

{\color{black} In our analysis, we use the gauge invariant variable in the spatially flat gauge where $\Psi=0$. This variable is defined as}
\begin{equation}\label{dphipsi}
  \delta\phi_{\Psi}\equiv\delta\phi+\frac{\dot{\phi}}{H}\left[1+2 \kappa_{4}^{2}f\left(1+\frac{\tilde{\rho}}{\sigma} \right)\right] \Psi.
\end{equation}
Using Eqs. \eqref{0iGperturbeddd}, \eqref{motionslow} and \eqref{Phi} to eliminate the remaining metric perturbations. In this gauge Eq. \eqref{perturbedmotionslow} can be written as
\begin{widetext}
{\small
\begin{equation}\label{aa}
 \ddot{\delta\phi_{\psi}}+3H\dot{\delta\phi_{\psi}}+\left[V_{,\phi\phi}-12 H^{2} f_{,\phi\phi} - \tilde{\kappa}_{\textrm{ eff}}^{2}\frac{V_{,\phi}\left(V_{,\phi}-6H^{2}f_{,\phi}\right)}{3H^{2}} \right] \delta\phi_{\psi}=0.
\end{equation}}
\end{widetext}
Introducing a new variable $v=a \delta\phi_{\Psi}$ which is called the Mukhanov-Sasaki variable, the perturbed equation of motion Eq. \eqref{aa} can be rewritten as
\begin{equation}\label{sasaki}
  v''-\frac{1}{\tau^{2}}\left(\nu^{2}-\frac{1}{4}\right)v=0,
\end{equation}
where the prime denotes the derivative with respect to the conformal time $\tau$ and the term $\nu$, to a first order, is given by
\begin{equation}\label{zsecond}
  \nu\simeq\frac{3}{2}+\epsilon -\eta +\frac{\zeta}{3}+2 \chi,
\end{equation}
where the slow-roll parameters are defined as follows
\begin{align}\label{epsilon}
  \epsilon&\equiv-\frac{\dot{H}}{H^{2}}\simeq\frac{1}{2\kappa_{4}^{2}}\left(\frac{V_{\phi}}{V}\right)^{2}C_{f,c}^{(a)},\\
   \eta&\equiv \frac{V_{,\phi\phi}}{3 H^{2}}\simeq \frac{1}{\kappa_{4}^{2}}\left(\frac{V_{,\phi\phi}}{V}\right)C_{f,c}^{(b)},\\
\zeta&\equiv 12 f_{,\phi\phi},\\ \label{zetaetchi}
\chi&\equiv  \frac{\tilde{\kappa}_{\textrm{ eff}}^{2}V_{,\phi}\left(V_{,\phi} -6 f_{,\phi}H^{2}\right)}{18H^{4}}\nonumber\\
&\simeq\dfrac{2\kappa_{4}^{2} (1-f_{,\phi}F_{1})\left[1+(1+2\kappa_{4}^{2}f)F_{2}\right]}{F_{1}^{2}\left[1+(1+2\kappa_{4}^{2}f) (1+2\kappa_{4}^{2}fF_{2})\right]}, 
\end{align}
where {\color{black} we have introduced corrections terms (respect to the standard 4d case) defined as}
 \begin{align}\label{correctionterms}
 C^{(a)}_{f,c}=& \frac{(1+2\kappa_{4}^{2}f)\left(1+\sqrt{1-U}\right)^{2}}{4\sqrt{1-U}}\left(1-2 f_{,\phi} F_{1}\right)\left(1-f_{,\phi} F_{1}\right),\\
  C^{(b)}_{f,c}= &\frac{(1+2\kappa_{4}^{2}f)(1+\sqrt{1-U})}{2}, \\
 F_{1}=&\frac{4\kappa_{\textrm{\textrm{ eff}}}^{2}V}{\left(1+\sqrt{1-U}\right)V_{,\phi}},\\
F_{2}= &(1+2\kappa_{4}^{2}f) \left[U- 4 f \kappa_{\textrm{ eff}}^{2} (1-\sqrt{1-U})\right].
\end{align}
These corrections terms depend on both the effect of the holographic cosmology (through $U$ terms) as well as the NMC (through $f$ terms). One can notice that at the low energy limit ($V\ll V_{max}$) and for $f\rightarrow 0$, the corrections terms reduce to one and the standard expressions of the slow roll parameters are recovered. These corrections terms can reproduce exactly the same form obtained for constant $f$ in \cite{Bargach:2019pst}.\par
 Finally we obtain the solution for \eqref{sasaki} as \cite{Riotto:2002yw}
\begin{equation}\label{deltaphi}
  v = \frac{aH}{\sqrt{2 k^{3}}} \left(\frac{k}{a H}\right)^{\frac{3}{2}-\nu}.
\end{equation}
The power spectrum for the scalar field perturbations is given by \cite{Bassett:2005xm}
\begin{equation}\label{pr}
  P_{\delta\phi}\equiv\frac{4 \pi k^{3}}{(2\pi)^{3}}\left|\frac{v}{a}\right|^{2}.
\end{equation}

We are now ready to calculate the spectral index of the power spectrum given by \cite{Bassett:2005xm}
\begin{equation}\label{nR}
  n_{r} -1\equiv \left(\frac{d \ln P_{\delta\phi}}{d \ln k}\right)_{k=aH}= 3-2\nu.
\end{equation}
In terms of slow-roll parameters and to a first order, the spectral index reads
\begin{equation}\label{nrSlow}
  n_{r} \simeq 1-2\epsilon+2\eta-4\chi.
\end{equation}

The power spectrum of the curvature perturbations {\color{black} in our model is} given by \cite{Bassett:2005xm}
\begin{align}\label{as}
  A_{s}^{2}&\equiv\frac{4}{25}P_{R}= \frac{4}{25} \frac{4 \pi k^{3}}{(2\pi)^{3}} |\mathcal{R}|^{2}\\
  &=\left(\frac{2H}{5\dot{\phi}\left[1+2 \kappa_{4}^{2}f\left(1+\frac{\tilde{\rho}}{\sigma} \right)\right]}\right)^{2} P_{\delta\phi},
  \end{align}
{\color{black} and can be rewritten in the slow-roll approximation as}
\begin{align}\label{asmod}
 A_{s}^{2}&= \frac{4}{25(2\pi)^{2}}\frac{H^{4}}{\dot{\phi}^{2}\left[1+2f\kappa_{4}^{2}(1+\frac{\tilde{\rho}}{\sigma})\right]^{2}}\nonumber\\&\simeq \frac{\kappa_{4}^{6} V^{3}}{75 \pi^{2} V_{\phi}^{2}} G_{f,c},
\end{align}
where the correction to the standard $4$-dimensions expression is given by
\begin{equation}\label{asmodi}
 G_{f,c}= \left(C_{f,c}^{(b)}\right)^{-3}\left[ (1-2 f_{,\phi}F_{1}) (1+2 \kappa_{4}^{2}f)\left(1+2 \kappa_{4}^{2}fF_{2}\right)\right]^{-2}.
\end{equation}
This correction term depends on the NMC and on the holographic cosmology effect. It reduces to one at low energy limit, i.e. $U\ll 1$ and for $f\rightarrow 0$.\par

\section{Higgs inflation model}\label{sec3}
In this section, we formulate a Higgs inflationary model from an holographic cosmology perspective as we have developed in the previous section. We will take into account observational constraints in order to check the viability of the model. In the Higgs inflationary model, the quadratic form of the NMC and the Higgs potential are respectively \cite{Rasanen:2018fom}
 \begin{equation}\label{f}
f (\phi) = \dfrac{1}{2\kappa_{4}^{2}}+\frac{\alpha_{0} \phi^{2}}{2},
\end{equation}
\begin{equation}\label{potential}
  V(\phi) =  \frac{\lambda}{4} \phi^{4},
\end{equation}
where $ \alpha_{0} $ is a coupling constant and $\lambda$ is the Higgs self-coupling.  Both quantities are dimensionless.\par

The inflationary regime refers to large values of the field. Inflation takes place when \mbox{$\alpha_{0}\kappa_{4}^{2} \phi^{2} \gg 1$}. Thus, the dimensionless parameter $U$, defined in Eq. \eqref{U}, is too small, i.e. $U\ll 1$.
Therefore, we can determine the range of values for the lower bound on the coupling constant by using the expression of the dimensionless parameter $U$. In this limit, we find
\begin{equation}
\alpha_{0} \gg \sqrt{\frac{2 c \lambda}{3}}\sim 2\times 10^{3},
\end{equation}
for $c=4\times10^{7}$ and $\lambda=0.13$. Those values are given in \cite{Bouabdallaoui:2016izz, Bargach:2019pst} and \cite{Rasanen:2018fom}, respectively. {\color{black} On the other hand, successful Higgs inflation is possible only if the Higgs field has a large NMC which allow to align the spectrum of primordial perturbations with observational constraints \cite{Bezrukov:2007ep}. Furthermore, while this large value of the NMC is in agreement with our result, it induces a unitarity violation at tree-level. Before concluding this section, we will come back to this issue.}\par
\begin{figure*}
  \centering
    \begin{tabular}{ccc}
    \subfloat[\label{fig1a}]{\includegraphics[scale=0.6]{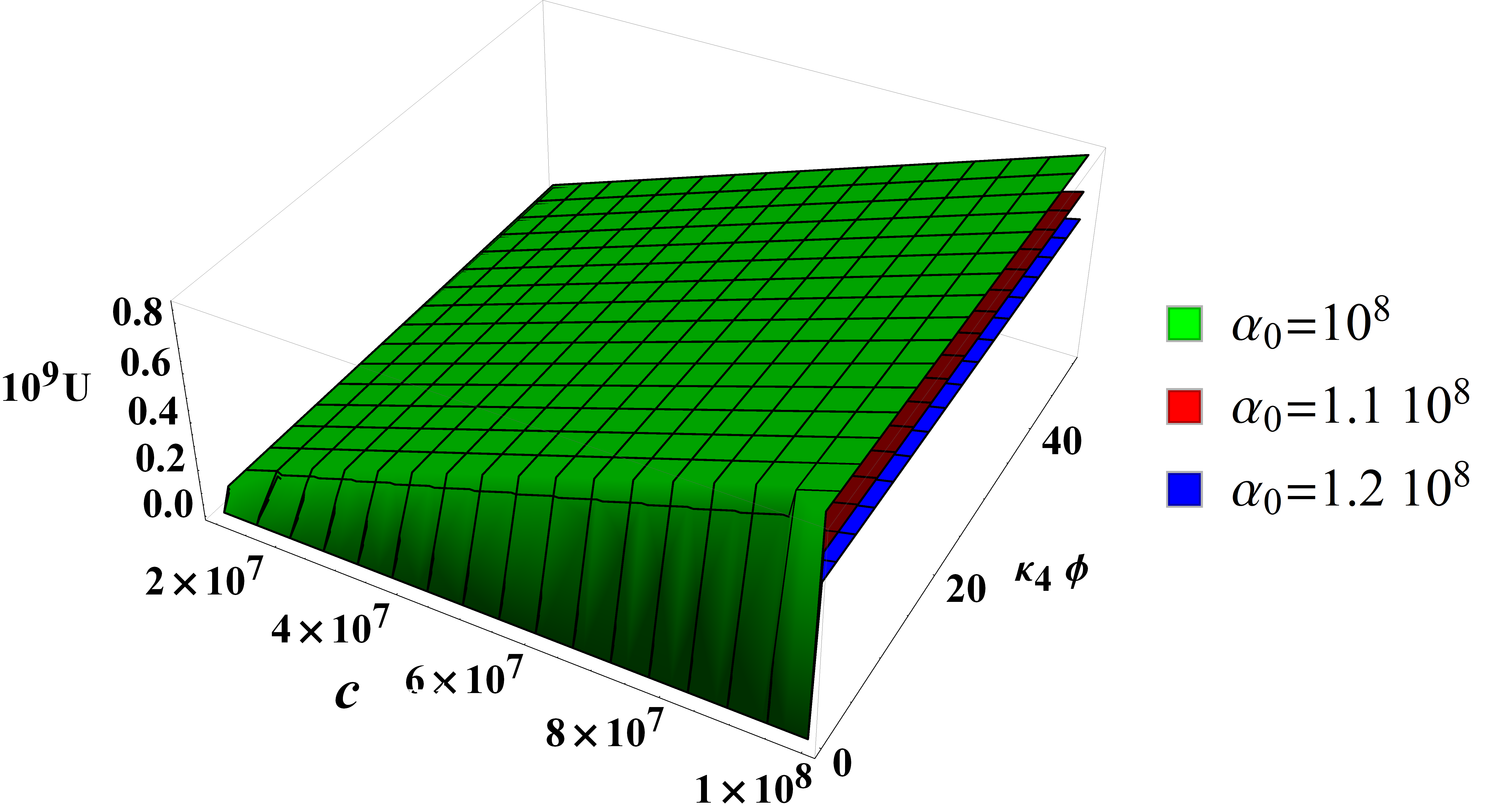}}& \qquad& \subfloat[\label{fig1b}]{\includegraphics[scale=0.6]{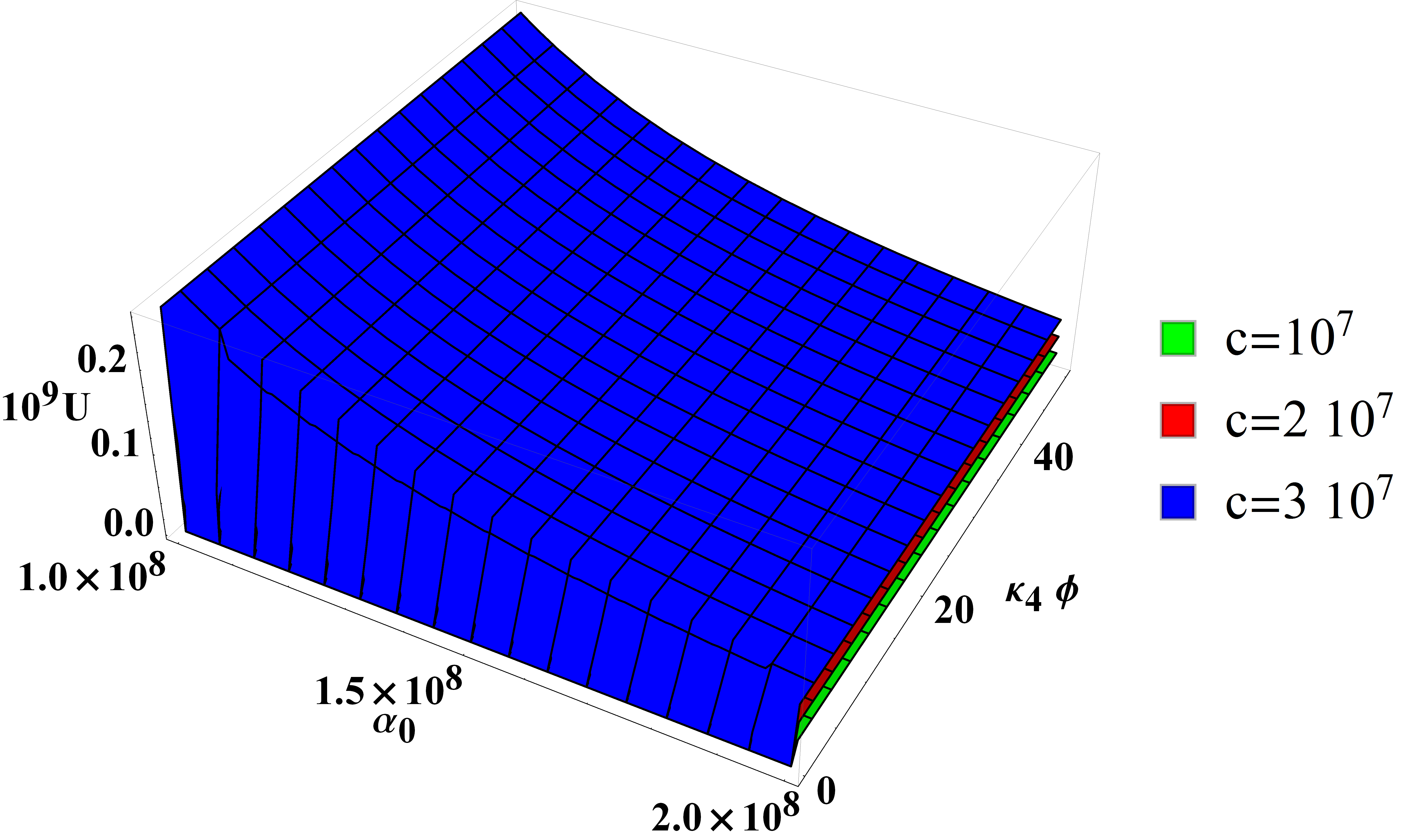}}\\
    \end{tabular}
   \caption[]{Evolution of the dimensionless parameter $U$ versus the conformal anomaly coefficient  $c$ and the scalar field $\phi$, for $\lambda=0.13$.}\label{figure1}
\end{figure*}

Fig .~\ref{fig1a} shows the variation of the dimensionless parameter $U$ versus the conformal anomaly coefficient $c$ and the scalar field $\phi$ for different values of the NMC constant $\alpha_{0}$. Fig .~\ref{fig1b} shows the variation of the dimensionless parameter $U$ versus the NMC constant $\alpha_{0}$ and the scalar field $\phi$ for different values of the conformal anomaly coefficient $c$. We conclude from these figures that a noticeable 
effect of the holographic cosmology and the NMC 
is usually at large values of the scalar field as we are looking for a small value of $U$ but still different from zero.\par   
The number of e-folds during inflation is given by
 \begin{equation}\label{efolds}
 N=  \int_{t_{i}}^{t_{f}}H dt,
 \end{equation}
 which in the slow-roll approximation can be written as
 \begin{equation}\label{efoldsslow}
  N\simeq-\int_{\phi_{i}}^{\phi_{f}}\frac{2\kappa_{\textrm{ eff}}^{2}\left(1-\sqrt{1-U}\right)}{U_{,\phi}+4f_{,\phi}\kappa_{\textrm{ eff}}^{2}\left(-2+U+2\sqrt{1-U}\right) }d\phi,
 \end{equation}
where $\phi_{i}$ being the value of the field at the crossing horizon and $\phi_{f}$ its value at the end of inflation.\par

In the large-field limit, using Eqs. \eqref{f} and \eqref{potential} the number of e-folds \eqref{efolds} reads
\begin{equation}\label{Nhc}
N=\frac{\left(c \lambda-3 \alpha_{0}^2\right) }{4 \alpha_{0} c \lambda} \log \left(\frac{\phi_{f}}{\phi_{i}}\right).
\end{equation}
Inflation stops at $\epsilon=1$ and Eq. \eqref{epsilon} together with Eqs. \eqref{f} and \eqref{potential}, implies
 \begin{align}\label{epsilonhc}
\nonumber \epsilon=&\frac{8 (2+\alpha_{0} \kappa_{4}^{2}\phi^{2})}{\kappa_{4}^{2}\phi^{2}} \left(1-\frac{2 \alpha_{0} \kappa_{4}^{2}\phi^{2}}{(2+\alpha_{0} \kappa_{4}^{2}\phi^{2}) (2-\frac{U}{2}+\frac{U^{2}}{4})} \right)\\
 &\times \left(1-\frac{ \alpha_{0} \kappa_{4}^{2}\phi^{2}}{(2+\alpha_{0} \kappa_{4}^{2}\phi^{2}) (2-\frac{U}{2}+\frac{U^{2}}{4})} \right).
 \end{align}

\begin{figure*}
  \centering
    \begin{tabular}{ccc}
    \subfloat[\label{fig3a} $c= 2\times 10^{7}$ ]{\includegraphics[scale=0.8]{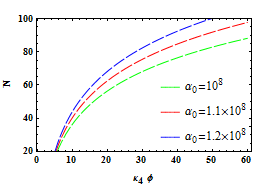}} & \qquad \qquad &
    \subfloat[\label{fig3b} $\alpha_{0}=10^{8}$ ]{\includegraphics[scale=0.8]{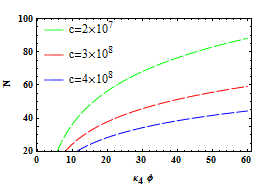}}\\
    \end{tabular}
   \caption[]{Evolution of $N$, the number of e-fold, versus  versus the scalar field $\phi$ for $\lambda=0.13$.}\label{figure3}
\end{figure*}

Fig .~\ref{figure3} shows the evolution of the number of e-folds $N$ against the Higgs field for a Higgs self-coupling $\lambda=0.13$. Where we plot in Fig .~\ref{fig3a} the evolution of $N$  against the Higgs field for different values of the NMC $\alpha_{0}$ while we fix the conformal anomaly coefficient to $c=2\times 10^{7}$. In Fig .~\ref{fig3b} we plot the evolution of $N$  against the Higgs field for different values of the conformal anomaly coefficient $c$ while we fix the NMC to $\alpha_{0}=10^{8}$. We conclude from these figures that we can find a range for the Higgs field for which the number of e-folds is in the appropriate range $50 < N < 70$.\par

We can as well derive the scalar perturbation, given,  Eq. \eqref{asmod}, and the scalar spectral index, see  Eq. \eqref{nR}, at the horizon crossing. They read
{\small
\begin{widetext}
\begin{align}
A_{s}^{2}=&\frac{\lambda  \kappa_{4}^{6}\phi^{6}}{4800 \pi^{2}}\frac{\left(1+(1+\alpha_{0} \kappa_{4}^{2}\phi^{2})\left[(2+\alpha_{0} \kappa_{4}^{2}\phi^{2})U-2(1+\alpha_{0} \kappa_{4}^{2}\phi^{2})(\frac{U}{2}-\frac{U^{2}}{4})\right]\right)^{-2}}{\left(2+\alpha_{0} \kappa_{4}^{2}\phi^{2}\right)^{5}\left(1-\frac{3U}{4}+\frac{3U^{2}}{8}\right)} \left(1-\frac{2\alpha_{0} \kappa_{4}^{2}\phi^{2}}{(2+\alpha_{0} \kappa_{4}^{2}\phi^{2})(2-\frac{U}{2}+\frac{U^{2}}{4})}\right)^{-2},\\
\nonumber n_{r}=&1- \frac{4 \left(2+\alpha_{0} \kappa_{4}^{2}\phi^{2}\right)}{\kappa_{4}^{2}\phi^{2}} \left[4 \left(1-\frac{2 \alpha_{0} \kappa_{4}^{2}\phi^{2}}{(2+\alpha_{0} \kappa_{4}^{2}\phi^{2}) (2-\frac{U}{2}+\frac{U^{2}}{4})} \right)\left(1-\frac{ \alpha_{0} \kappa_{4}^{2}\phi^{2}}{(2+\alpha_{0} \kappa_{4}^{2}\phi^{2}) (2-\frac{U}{2}+\frac{U^{2}}{4})} \right)+3\left(2-\frac{U}{2}+\frac{U^{2}}{4}\right)\right]
-24 \alpha_{0}\\
&-\frac{16 (2+\alpha_{0} \kappa_{4}^{2}\phi^{2}) (2-\frac{U}{2}+\frac{U^{2}}{4}) \left[(2+\alpha_{0} \kappa_{4}^{2}\phi^{2})U-2(1+\alpha_{0} \kappa_{4}^{2}\phi^{2})(\frac{U}{2}-\frac{U^{2}}{4})\right]}{\kappa_{4}^{2}\phi^{2}\left(1+(1+\alpha_{0} \kappa_{4}^{2}\phi^{2})\left[(2+\alpha_{0} \kappa_{4}^{2}\phi^{2})U-2(1+\alpha_{0} \kappa_{4}^{2}\phi^{2})(\frac{U}{2}-\frac{U^{2}}{4})\right]\right)}\left(1-\frac{ \alpha_{0} \kappa_{4}^{2}\phi^{2}}{(2+\alpha_{0} \kappa_{4}^{2}\phi^{2}) (2-\frac{U}{2}+\frac{U^{2}}{4})} \right),
\end{align}
\end{widetext}
}
respectively.\par

\begin{figure*}
  \centering
    \begin{tabular}{ccc}
    \subfloat[\label{fig4a} ]{\includegraphics[scale=0.6]{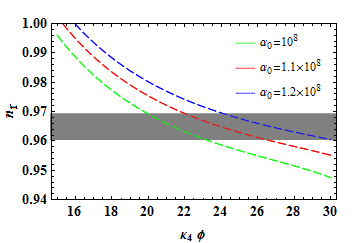}} & \qquad &  
    \subfloat[\label{fig4b} ]{\includegraphics[scale=0.6]{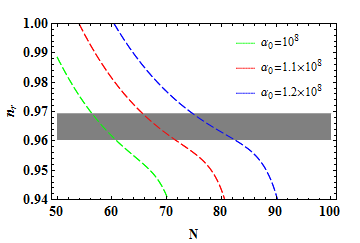}}
    \\
    \end{tabular}
   \caption[]{Evolution of $n_{r}$ versus the number of e-folds $N$ for  $\lambda=0.13$ and $c= 2\times 10^{7}$. Also, we show the $1 \sigma$ bound on $n_{r}$ from Planck data \cite{Akrami:2018odb} with the gray horizontal region.}\label{figure4}
\end{figure*}

We show in Fig .~\ref{figure4} the evolution of $n_{r}$  for $\lambda=0.13$ and $c=2\times 10^{7}$, together with the bound on $n_{r}$ from Planck data \cite{Akrami:2018odb}. In Figs .~\ref{fig4a} and \ref{fig4b} we plot $n_{r}$ versus the Higgs field $\phi$ and $n_{r}$ versus the number of e-folds $N$, respectively. Both plots are made for different values of the NMC $\alpha_{0}$. We notice that the predictions of $n_{r}$ are consistent with Planck data.\par

\begin{figure}
  \centering
{\includegraphics[scale=0.54]{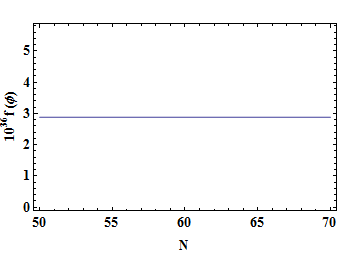}} 
   \caption[]{Evolution of $f(\phi)$ versus the number of e-folds $N$ for $\lambda=0.13$, $\alpha_{0}=10^{8}$ and $c= 4\times 10^{7}$.}\label{figuref}
\end{figure}

We next evaluate the tensor-to-scalar ratio to further check the viability of our model. Using Eqs. \eqref{f}, \eqref{epsilonhc} and \eqref{Nhc}, we plot in Fig.~\ref{figuref} the variation of the NMC function $f(\phi)$ against the number of e-folds in the range $50<N<70$ for $\lambda=0.13$, $\alpha_{0}=10^{8}$ and $c= 4\times 10^{7}$. This figure shows that $f(\phi)$ does not change much. In this case, we can use the expression of tensor perturbations for a constant induced gravity correction at the Hubble crossing which reads \cite{BouhmadiLopez:2004ax}
\begin{equation}\label{At}
  A_{T}^{2}= \frac{2 \kappa_{4}^{2}}{25 } \left(\frac{H^{2}}{2 \pi}\right)^{2} F^{2}\left(\frac{H}{\mu}\right),
\end{equation}
where $\mu= \kappa_{4}^{2}/\kappa_{5}^{2} (1-\gamma)$, $\gamma$ is the induced gravity constant and the function $F(x)$ is given by
\begin{equation}
F^{-2}(x)=\gamma+(1-\gamma)\left[\sqrt{1+x^{2}}-x^{2}\arcsinh  \frac{1}{x}\right].
\end{equation}

\begin{figure}
  \centering
    \begin{tabular}{c}
    \label{fig5a}{\includegraphics[scale=0.6]{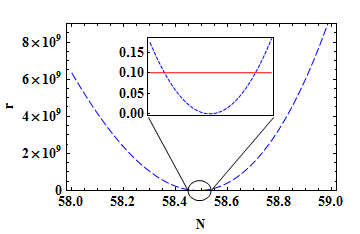}} \\
    \end{tabular}
   \caption[]{Evolution of $r$ versus the number of e-folds $N$ for  $\lambda=0.13$ and $c= 2\times 10^{7}$. We show also a subplot where we have zoomed the wanted range for which our predictions lies in the bound imposed by Planck data. The horizontal red line indicates the upper bound for the tensor-to-scalar ratio imposed by Planck data.}\label{figure5}
\end{figure}

In our model and within the slow-roll approximation, we find
{\small
\begin{widetext}
\begin{equation}\label{Atmod}
   A_{T}^{2}\simeq \frac{\lambda \kappa_{4}^{4}\phi^{4}}{300 \pi^{2}(1+\alpha_{0} \kappa_{4}^{2}\phi^{2})}\left[1-\sqrt{1+4 \alpha_{0} \kappa_{4}^{2}\phi^{2} \left(\frac{U}{2}-\frac{U^{2}}{4}\right)}+4(2+\alpha_{0} \kappa_{4}^{2}\phi^{2})\left(\frac{U}{2}-\frac{U^{2}}{4}\right) \arcsinh\left(\sqrt{4(2+\alpha_{0} \kappa_{4}^{2}\phi^{2})\left(\frac{U}{2}-\frac{U^{2}}{4}\right)}\right)^{-1}\right]^{-1}.
\end{equation}
\end{widetext}}
Finally, the ratio between the amplitudes of tensor and scalar perturbations is given by
\begin{equation}\label{r}
  r\equiv \frac{A_{T}^{2}}{A_{S}^{2}}.
\end{equation}
In Fig .~\ref{figure5}, we plot the evolution of $r$ versus the number of e-folds $N$ for  $\lambda=0.13$ and $c= 2\times 10^{7}$. It is important to notice that the tensor-to-scalar ratio lies within the
bound imposed by Planck data in an extremely small range of $N$ which is around $\sim 58.48$. \par
\begin{figure}
  \centering
    \begin{tabular}{c}
   \label{fig6a} {\includegraphics[scale=0.6]{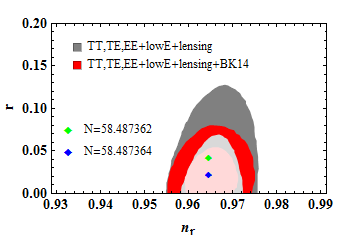}} \\
    \end{tabular}
   \caption[]{Plot of the tensor-to-scalar ratio $r$ against the scalar spectral index $n_{s}$ for $\alpha_{0}=10^{8}$, $\lambda=0.13$ and $c=2\times 10^{7}$. The marginalized joint $68$\% and $95$\% confidence
level contours $(n_{r}, r)$ using Planck alone and in combination with BK14 data.}\label{figure6}
\end{figure}

Finally, to compare the consistency between our theoretical predictions for a NMC Higgs field and observations we plot the ($n_{r},r$) plane in Fig .~\ref{figure6}, where we show constraints from the Planck TT, TE, EE+lowE+lensing data (gray contour) and Planck TT,
TE, EE+lowE+lensing+BK14 data (red contour). We can see that our predicted parameters lie inside the 95\% C. L. of the Planck data for the two values of the selected number of e-folds for $\alpha_{0}=10^{8}$, $\lambda=0.13$ and $c=2\times 10^{7}$.\par

{\color{black} The large NMC to gravity, which ensures  successful Higgs inflation, violates at tree-level unitarity at a scale corresponding to inflation \citep{Burgess:2009ea,Lerner:2009na,Barbon:2009ya,Hertzberg:2010dc,McDonald:2016cdh,Nozari:2018fve,Rubio:2018ogq}. This means that the theory as it stands is incomplete.  The unitarity cutoff scale of the theory has been extensively studied in several works due to the importance of  Higgs inflation, see for example \cite{Lerner:2011it} for some proposed solutions to this problem. \par
Let us check the tree-level unitarity violation of the effective field theory for the model under study in the large field regime. Since the Higgs field is NMC to the Ricci scalar of the induced metric, standard 4D result is expected to remain true for our model also i.e. the effective field theory holds only at energy scales not higher than
\begin{equation}
\phi \sim \frac{1}{\kappa_{4} \alpha_{0}}.
\end{equation}
In standard inflation, the inflationary phase exits at ($\epsilon=1$) i.e.
 \begin{equation}
\phi_{f} \sim \frac{1}{\kappa_{4} \sqrt{\alpha_{0}}}.
\end{equation}
This means that the tree-level analysis is invalid due to unitarity violation at the scale ($1/\kappa_{4} \alpha_{0}$). Due to the complexity of the expression of $\epsilon$ in our model (Eq. \eqref{epsilonhc}), we plot in Fig. \ref{figure7} the Higgs field values at the end of inflation with respect to the conformal anomaly coefficient $c$ for different values of the NMC constant $\alpha_{0}$ and for $\lambda=0.13$, together with the upper bound of unitarity given by $\kappa_{4} \phi_{f}= 1/\alpha_{0}$. We see from this figure that $\kappa_{4} \phi_{f}$ is significantly larger than the upper bound which we present by horizontal regions for different values of the NMC constant. We notice that for $c<10^{8}$, $\kappa_{4} \phi_{f}$ are almost constant whatever is the value of $\alpha_{0}$. {\color{black} On the other hand, we can see from Fig. \ref{figure1} that the holographic duality (through the anomaly coefficient $c$) has a noticeable effect on the standard cosmology dynamic only for large values of the anomaly coefficient ($c$ $\sim 10^{7}$). This justify the choice of the large values of the $c$-parameter in the paper for which the imprints of holographic cosmology appears clearly. This range is also appropriate} for the inflationary parameters in the context of the holographic duality \cite{Lidsey:2005nt, Bouabdallaoui:2016izz, Bargach:2019pst}. While for $c>10^{8}$, the behaviour of $\kappa_{4} \phi_{f}$ changes but still well above the unitarity bound for all values of the NMC constant. The minimum in this figure shows the maximal value of the conformal anomaly coefficient corresponding to $U=1$ for example for $\alpha_{0}=10^{6}$, $c_{\textrm{max}}\sim 10^{13}$.\par 
\begin{figure}
  \centering
    \begin{tabular}{c}
   \label{fig7} {\includegraphics[scale=0.54]{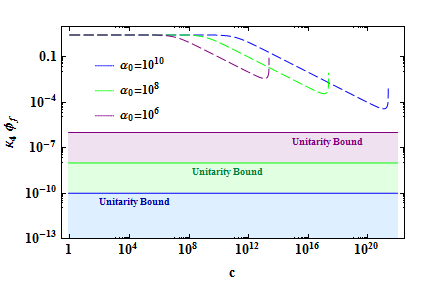}} \\
    \end{tabular}
   \caption[]{{\color{black}Evolution of $\phi_{f}$ versus the conformal anomaly coefficient $c$ for $\lambda=0.13$. We show the unitarity bounds with the horizontal regions for different values of the NMC constant.}}\label{figure7}
\end{figure}

Unfortunately, a typical inflation scale will be higher than our unitarity bound. So, the scale of the tree-level unitarity is violated. {\color{black}This result is not influenced by the values of the conformal anomaly coefficient}. However, as mentioned in \cite{Bezrukov:2010jz,Ferrara:2010in}, this result does not necessarily spoil the self-consistency of the Higgs inflationary scenario.\par
}

\section{Conclusions} \label{sec4}
In this paper, we have studied an inflationary scenario where the field is NMC with gravity in the framework of holographic cosmology from a braneworld point of view in the Jordan frame. We carried our analysis with a slow-roll approach.\par
Furthermore, we have analysed  the model from a  background and a perturbative analyssis obtaining the representative parameters as shown in sections \ref{sec1} and \ref{sec2}. The holographic nature of the setup together with the effect of the NMC is manifest through the existence of corrections terms for the standard background and perturbative parameters.\par

As an application of the model we have developed a Higgs inflationary model, where we have assumed a Higgs field NMC to Ricci scalar. In a large field limit, a quartic potential with a self-coupling $\lambda=0.13$, as fixed by observations \cite{Rasanen:2018fom}, and a NMC $\alpha_{0}=10^{8}$ with a conformal anomaly coefficient $c=2\times 10^{7}$ lies extremely well with observations made by Planck 2018 for a number of e-folds $N\sim 58.48$, where the value of the scalar spectral index {\color{black}and the tensor-to-scalar} ratio turned out to be $n_{r}=0.965$ and $r=0.021$, {\color{black}respecively}. However, as soon as we move from $N\sim 58.48$, the model is ruled out by the current data. \par

{\color{black}As we have already mentioned, the Higgs field is the unique scalar field in standard model of particle physics, i.e. it is the most economical model when Higgs itself can drive inflation. In this paper, we deal with an inflationary model driven by the Higgs confined on a brane which could be seen as one of the economical models and also it is in good agreement with observations as noticed in the previous section (see Fig. \ref{figure6}). A complemantery study of the unitarity violation at tree level is needed to determine the validity of our analysis.  We will leave this question for future work.\par}

\section*{Acknowledgment }
The work of MBL is supported by the Basque Foundation of Science Ikerbasque. She also would like to acknowledge the partial support from the Basque government Grant No. IT956-16 (Spain) and from the project FIS2017-85076-P (MINECO/AEI/FEDER, UE).

\appendix
\section{Weyl tensor for a brane-wrold model with a general induced gravity term} \label{appendix}
In Section III, we showed that the trace-free projection of the $5$-dimensions Weyl tensor, $E_{\mu}^{\nu}$, cannot be neglected at the perturbative level since it encodes the effects of the bulk gravitational field on the brane. In this appendix, we show that instead of neglecting $\delta E_{\mu}^{\nu}$, we parametrise this tensor as an effective fluid (see Eq. \eqref{deltaE}) and then we find a relationship between its components. Our objective in this appendix is to check if a similar result as the one in \cite{Koyama:2005kd} for the Dvali-Gabadadze-Porrati (DGP) model, can be obtained for a branewrold model with a general  induced gravity term.\par

Our starting point is the Bianchi identity
\begin{equation}\label{bianchi}
 \nabla_{\mu} G^{\nu}_{\mu}=0, 
 \end{equation}
and the equation of conservation of the energy-momentum {\color{black}tensor}, Eq. \eqref{continuite}. From Eq. \eqref{Gmodifie}, one obtain \cite{Koyama:2005kd}
 \begin{equation}\label{EetPi}
 \nabla_{\mu} E^{\nu}_{\mu}= \kappa_{5}^{4} \nabla_{\mu} \Pi^{\nu}_{\mu}.
 \end{equation}
The $(t, t)$ and the $(0, i)$ components of the perturbed $4$-dimensions field equations Eq. \eqref{EetPi} are given, respectively, by \cite{Koyama:2005kd}
  \begin{align}
 \delta \dot{\rho_{E}}+4 H \delta \rho_{E} +\Delta \delta q_{E}&=0, \label{nu=0}\\
 \delta \dot{q_{E}}+4 H \delta q_{E} +\frac{1}{3} \left(\delta \rho_{E} +2 \Delta \delta\pi_{E} \right)&=
  -\frac{ (\tilde{\rho}+\tilde{p})}{\sigma} \tilde{\rho} \Delta,\label{nu=i}
 \end{align}
where $ \tilde{\rho} \Delta =\left(\delta \tilde{\rho} -3H \delta \tilde{q}\right).$
Using the quasi-static approximation, we can neglect time-derivative terms relative to gradient ones, 
we get from Eq. \eqref{nu=0} 
\begin{equation}\label{deltaqe=0}
\delta q_{E}=0.
\end{equation}
Furthermore the Bianchi identity suggests that there exists a family of solutions characterized by \cite{Koyama:2005kd} 
\begin{equation}
\delta \rho_{E}=C\left(\sigma \tilde{\rho} \tilde{p}\right) k^{2} \delta\pi_{E}.
\end{equation}
In order to obtain the correct form of the function $C\left(\sigma \tilde{\rho} \tilde{p}\right)$ some global $5$-dimensional perturbation analysis of the Weyl fluid for a RSII with IG model will be
necessary to solve this issue.\par
In this paper, the result \eqref{deltaqe=0} is enough to continue our investigation. Thus, we have checked this result for a branewrold model with a general induced gravity term which has an anti de-Sitter space-time bulk geometry.

\end{document}